\documentstyle[preprint,aps]{revtex}
\newcommand{\be}{\begin{equation}}
\newcommand{\ee}{\end{equation}}
\newcommand{\bea}{\begin{eqnarray}}
\newcommand{\eea}{\end{eqnarray}}

\newcommand{\p}{\partial}

\tightenlines 

\begin{document}

\draft
\preprint{\small   }

\title{Diffeomorphism invariant $SU(N)$ gauge theories}

\author{Viqar Husain\footnote{
Email: husain@physics.ubc.ca}}

\address{\baselineskip=1.4em Department of Physics and Astronomy,\\
University of British Columbia,\\
6224 Agricultural Road \\
Vancouver, BC V6T1Z1, Canada}
\maketitle 
\medskip

\begin{abstract}

We describe a class of diffeomorphism invariant $SU(N)$ gauge theories
in $N^2$ dimensions, together with some matter couplings. These theories 
have $(N^2-3)(N^2-1)$ local degrees of freedom, and have the unusual 
feature that the constraint associated with time reparametrizations is
identically satisfied. A related class of $SU(N)$ theories in $N^2-1$
dimensions has the constraint algebra of general relativity, but has  
more degrees of freedom. Non-perturbative quantization of the first type 
of theory via $SU(N)$ spin networks is briefly outlined.

\end{abstract}

\bigskip


Generally covariant theories have the well known feature that their
Hamiltonia are linear combinations of constraints in the canonical 
formulation. The Poisson algebra of constraints in the Hamiltonian  
theory normally reflects all the spacetime symmetries of the covariant 
theory. This is the case in theories such as general relativity and 
supergravity, in which the spacetime metric is non-degenerate in general. 
The constraint associated with time reparametrization plays a dual 
role: it may be viewed as generating both unphysical gauge transformations 
on phase space variables because of its origin, and  physical 
``time-evolution''  because it gives the rule for evolving initial 
data from one spatial surface to the next. A true non-vanishing 
Hamiltonian appears only after a time gauge fixing is made; physical 
evolution is then unambiguous, but is with respect to the gauge choice. 

The normal expectation is that all local invariances of an action are 
manifested in the Hamiltonian theory via distinct first class constraints. 
However there are two known exceptions. 

The first is a theory with general coordinate and $SU(2)$ Yang-Mills 
invariance in four spacetime dimensions\cite{HK}. The fields are 
spacetime dreibeins $e_a^i$ and gauge fields $A_a^i$, where $a=0,\cdots,3$ 
is the spacetime index and $i=1,\cdots,3$ is the $SU(2)$  index. 
The fields are therefore in the adjoint representation. The action is 
\be 
      S_4[e,A] = \int \eta_{ijk}\ e^i\wedge e^j\wedge F^k(A), 
\label{su(2)}
\ee
where $F(A)=dA+[A,A]$ is the curvature of $A_a^i$, and 
$\eta_{ijk}$ is the $su(2)$ anti-symmetric tensor. This theory has been 
analyzed in detail in \cite{HK}. The phase space coordinates are exactly 
those of $SU(2)$ Yang-Mills theory. However, the dynamics is (obviously) 
different: the Hamiltonian turns out to be a linear combination of the 
Gauss law and spatial diffeomorphism constraint. The surprising feature 
of this theory is that {\it the constraint corresponding to time 
reparametrization invariance vanishes identically}.  

The second theory is a generalization of Chern-Simons theory to all odd 
dimensions higher than three. The action is again invariant under general
coordinate and Yang-Mills gauge transformations, but unlike (\ref{su(2)}), 
is a functional of a single gauge field \cite{Henn1,Henn2}. The Lagrangian 
${\cal L}$ for the $(2n+1)$-dimensional theory is defined via 
\be 
  d{\cal L}^{2n+1} = g_{i_1\cdots i_{n+1}} F^{i_1} \wedge \cdots 
\wedge F^{i_{n+1}},
\label{cs}
\ee
where now the gauge group may be $SU(N)$ rather than just $SU(2)$ as in 
(\ref{su(2)}), and $g_{i_1\cdots i_{n+1}}$ is a symmetric Lie algebra 
tensor of rank $n+1$ constructed from the structure constants. For $n>2$ 
these theories have local degrees of freedom. The Hamiltonian-Dirac analysis 
reveals the feature  that the time-reparametrization constraint arises 
as a linear combination of internal Yang-Mills  transformations and 
spatial diffeomorphisms: {\it there is no independent time diffeomorphism 
constraint} \cite{Henn1}. 

One of the main differences between these two theories at the Hamiltonian 
level is that in the first there is {\it no} combination of first class 
constraints that gives the time diffeomorphism constraint, while in the 
second this is the case. Furthermore, the phase space coordinates themselves 
are different: the coordinates for the second theory are spatial components 
of the gauge field as is usual for Chern-Simons theory,  rather than 
electric and connection fields as in the first theory.
Finally, while (\ref{cs}) is defined in all odd dimensions, it appears 
at first sight that (\ref{su(2)}) can be defined only in four dimensions, 
and only for $SU(2)$. Indeed, if the group is taken to be SO(3,1), 
(\ref{su(2)}) is an action for general relativity. For groups with 
more than three generators (in four spacetime dimensions), the time
diffeomorphism constraint is always present in the Hamiltonian formulation. 

In this note we point out that there are in fact higher dimensional
theories of the general form (\ref{su(2)}). We will see that this type 
of action, with the vanishing Hamiltonian constraint property, may be 
written down only for spacetime dimensions which are specific functions 
of the number of generators of the internal group.  The  
vanishing Hamiltonian constraint means that ``dynamics'' 
effectively amounts to the kinematics of general relativity in the 
respective dimensions. This makes the theories integrable quantum 
mechanically, as is the case \cite{ash} for the $SU(2)$ theory 
(\ref{su(2)}). Finally, we discuss possible matter couplings, and point 
out that there is another class of $SU(N)$  gauge theory in $N^2-1$ 
dimensions which has a canonical constraint algebra {\it identical} to 
the  4-dimensional general relativity constraint algebra in the Ashtekar 
formulation. This theory, however, has more degrees of freedom than 
general relativity in the same dimension. This demonstrates that 
identical constraint algebras need not imply identical theories 
classically or quantum mechanically.  

The key input responsible for the absence of the Hamiltonian
constraint in the theory (\ref{su(2)}) is that the 4-dimensional
action is a functional of drei-beins $e_a^i$. This means that the
spacetime metric $g_{ab}= e_a^ie_b^j\delta_{ij}$ is degenerate with 
signature $(0+++)$. One can attempt to formulate similar higher 
dimensional theories with degenerate metrics. For example, for a 
5-dimensional theory we would like to have a degenerate covariant 
metric of signature $(0++++)$, and might try $SO(4)$ (or $SO(3,1)$ etc.) 
as the internal gauge group.  However, this does not work because there 
is no way to form a gauge scalar lagrangian density of the desired type: 
the anti-symmetric tensor on the group is the Levi-Civita tensor 
$\eta^{ijkl}$, whereas the would be 5-form lagrangian density is  
\be 
e^i\wedge e^j\wedge e^k\wedge F^{lm}.
\ee 
Thus tracing of internal indices, for group $SO(4)$, isn't possible. 
The same problem occurs for all higher dimensions with $SO(N)$ groups.

For $SU(N)$ groups, however, it is possible to write down higher
dimensional actions of the form (\ref{su(2)}) because the gauge field
carries one internal index rather than an antisymmetrized pair as for
$SO(N)$.  But  now the requirement that the spacetime metric be
degenerate is possible only in certain dimensions. To see this
consider $SU(3)$ as an example: the field $e_a^i$ ($i=1,\cdots, 8$)
gives a degenerate metric via $g_{ab} = e_a^ie_b^j k_{ij}$ (where
$k_{ij}$ is the Cartan metric), only if the spacetime dimension is
nine. 

More generally it is possible to write down degenerate metric theories 
of this type with internal gauge group $SU(N)$ only if the spacetime 
dimension is $N^2$. The action is
\be 
   S^{N^2} = \int_M \eta_{i_1\cdots i_{N^2-1}} 
e^{i_1}\wedge \cdots \wedge e^{i_{N^2-2}}\wedge F^{i_{N^2 -1}}(A), 
\label{N2}
\ee
where 
\be 
\eta_{i_1\cdots i_{N^2-1}} = f^{k_0}_{\ {i_1}{k_1}}f^{k_1}_{\ {i_2}{k_2}}
f^{k_2}_{\ {i_3}{k_3}}\cdots f^{k_{N^2-2}}_{\ i_{N^2-1}{k_0}}    
\label{gcontr}
\ee
is a tensor of rank $N^2-1$ on the Lie algebra. This tensor has 
an antisymmetric part proportional to the internal Levi-Civita tensor.  
 ($i_1,\cdots ,i_{N^2-1}$ are $su(N)$ indices, and $f^i_{\ jk}$ 
are the structure constants). Like (\ref{su(2)}), this action is invariant 
under all general coordinate transformations, and under the Yang-Mills 
gauge transformations 
$\delta_\Lambda e^i = f^i_{\ jk} e^j\Lambda^k$ and $
\delta_\Lambda A^i = D\Lambda^i$, where $D$ is the covariant
derivative of $A$.

To study the dynamics further let us consider the Hamiltonian formulation 
of the action (\ref{N2}). Assuming that the $N^2$ dimensional manifold is 
$M=\Sigma\times R$ where $\Sigma$ is a compact space, choose coordinates 
$(x^0,x^a)$ on $M$ such that $x^0\in R$, and the $t=$ constant 
surfaces are slices with the topology of $\Sigma$. Let 
\be
\tilde{\epsilon}^{a_1\cdots a_{N^2-1}} = 
\tilde{\epsilon}^{0a_1\cdots a_{N^2-1}},  
\ee
be the Levi-Civita tensor density on $\Sigma$, where the right hand side 
is the same density on $M$, and 
\be 
  \tilde{e} = {1\over (N^2-1)!}\ \eta_{i_1\cdots i_{N^2-1}} \ 
\tilde{\epsilon}^{a_1\cdots a_{N^2-1}}\ 
\left( e^{i_1}_{a_1}\cdots e^{i_{N^2-1}}_{a_{N^2-1}} \right)
\ee
be the determinant of the fields $e_a^i$. Assuming that this 
``spatial'' determinant does not vanish, we can define the dual 
fields 
\be 
 e^a_i = {1\over \tilde{e}}\ \eta_{ii_1 \cdots i_{N^2-2}}\ 
\tilde{\epsilon}^{aa_1\cdots a_{N^2-2}}\ 
 \left( e^{i_1}_{a_1} \cdots e^{i_{N^2-2}}_{a_{N^2-2}} \right),
\ee
which satisfy the expected relations $e_a^i e^b_i = \delta_a^b$ and 
$e^a_i e_a^j = \delta_i^j$. With the additional definitions 
\be 
N^a = e_0^i e_i^a
\ee
and 
\be 
\Lambda^i = A_0^i - N^aA_a^i,
\ee
the action (\ref{N2}) becomes 
\be 
   S_{N^2} = \int_R dx^0 \int_\Sigma d^{N^2-1}x 
\left[ \tilde{e}^a_iA_a^i - N^a\tilde{H}_a - \Lambda^i\tilde{G}_i \right],
\ee
where 
\be
\tilde{G}_i= \p_a\tilde{e}^a_i + f_i^{\ jk}A_a^j \tilde{e}^a_k,
\label{gauss}
\ee
\be 
\tilde{H}_a = F_{ab}^i - A_a^i\tilde{G}_i.
\label{diff}
\ee
This is the desired Hamiltonian action. The indices $a$ are now 
associated with $\Sigma$ and the phase space variables are the pairs
$(A_a^i,\tilde{e}^a_i)$. Varying the action with respect to the
$\Lambda^i$ and $N^a$ gives the constraints
\be 
  \tilde{G}_i = 0 \ \ \ \ {\rm and}\ \ \ \  \tilde{H}_a = 0.  
\ee
The Hamiltonian is a linear combination of constraints as expected for a 
generally covariant theory. However, as is manifest, there is no constraint 
corresponding to time reparametrization invariance. This feature is also 
present in the 4-dimensional theory studied in Ref. \cite{HK}. The two 
sets of constraints (\ref{gauss}) and (\ref{diff}) are first class and, 
as expected, generate respectively 
Yang-Mills and spatial diffeormorphisms on the phase space variables. 
These are the kinematical constraints in the Ashtekar canonical variables
\cite{ash2}.

Since there are $(N^2-1)^2$ configuration degrees of freedom $A_a^i$, 
and $2(N^2-1)$ first class constraints per space point, the theory has
\be 
(N^2-1)(N^2-3)
\ee 
local degrees of freedom. This is significantly more than the $N^2(N^2-3)/2$ 
of local degrees of freedom of general relativity in $N^2$ spacetime 
dimensions.
 
The result that the Hamiltonian constraint is identically zero in this 
theory may be understood from the covariant point of view. Recall that 
the Hamiltonian constraint of general relativity arises as the projection 
of the field equations along a timelike direction. In the present case the 
vector density 
\be 
    \tilde{u}^a = {1\over (N^2-1)!}\  \eta_{i_1\cdots i_{N^2-1}}  
      \ \tilde{\epsilon}^{aa_1\cdots a_{N^2-1}}
  \left( e^{i_1}_{a_1} \cdots e^{i_{N^2-1}}_{a_{N^2-1}} \right)
\label{u}
\ee 
satisfies $\tilde{u}^ae_a^i = 0$, and defines a special ``timelike''
direction on $M$. 

We will now show that the projection of the covariant 
field equations along this direction vanishes identically. Varying  
(\ref{N2}) with respect to $e_a^i$ and projecting along $\tilde{u}^a$ 
gives 
\be 
  \eta_{i_1\cdots i_{N^2-1}}\ e^{i_2}\wedge \cdots \wedge e^{i_{N^2-2}}
\wedge \tilde{F}^{i_{N^2-1}} = 0,  
\label{eom1}
\ee 
where $\tilde{F}^i_a = F^i_{ab}\tilde{u}^b$. 
Now because $\tilde{F}^i_a \tilde{u}^a = 0$, we can 
write 
\be 
\tilde{F}^i_a = \tilde{F}^i_{\ j} e_a^j. 
\ee 
Finally, with $\tilde{F}_{i_1\cdots i_{N^2-2} j} 
:= \eta_{i_1\cdots i_{N^2-1}} \tilde{F}^{i_{N^2-1}}_{\ j}$, 
eqn. (\ref{eom1}) becomes 
\be 
\left( e^{i_2}\wedge \cdots \wedge e^{i_{N^2-2}}
\wedge e^j \right) \tilde{F}_{i_1\cdots i_{N^2-2} j} = 0.
\label{eom11}
\ee
Now $\tilde{F}_{i_1\cdots i_{N^2-2} j}$ is antisymmetric in the 
first $N^2-2$ indices by definition. However, it must also be symmetric 
in the last $N^2-2$ indices for Eqn. (\ref{eom11}) to hold. This means 
that $ \tilde{F}_{i_1\cdots i_{N^2-2} j}\equiv 0$, and hence 
$F^i_{ab}\tilde{u}^b \equiv 0.$  Thus, the projection of this equation 
of motion along $\tilde{u}$  vanishes identically. The same result holds 
for the other equation of motion 
\be 
     \eta_{i_1\cdots i_{N^2-1}} 
D\wedge e^{i_1}\wedge \cdots \wedge e^{i_{N^2-2}} = 0.          
\ee
The vanishing of these projections is the covariant reason that the 
Hamiltonian constraint is absent, or more precisely, identically 
satisfied, in the canonical theory.
 
The 4-dimensional theory \cite{HK} has been completely quantized via  
a ``lattice'' method \cite{ash}. In this approach, holonomies of the 
connection in various representations (``colors'') of $SU(2)$ are 
associated with the edges of a graph embedded in $\Sigma$.  From this 
collection of holonmies, Gauss law invariant states are constructed by 
forming gauge scalars by summing over group indices using ``intertwining'' 
matrices ($6-j$, $9-j$, symbols etc.), which are associated with the 
nodes of the graph. This means that the colors which meet at a node 
must be compatible with the usual angular momentum addition rules. 
This procedure gives the so called spin-network states 
\cite{lsspin,baez}.  Thus, the links of the graph carry 
colors, and the nodes carry group index contraction information. Finally, 
spatial diffeomorphism invariant states are formed by ``summing''
over the diffeomorphism group. In the final picture, this effectively 
amounts to the labelling of physical states by quantities associated 
with the equivalence classes of graphs under spatial diffeomorphisms. 
The diffeomorphism invariant information labelling graph states 
consists of group representations associated with the edges, and 
intertwining matrices associated with the nodes of a graph. This is 
what is expected intuitvely: diffemorphisms cannot ``break up, recolor, 
and resum'' the edges of graphs, or alter the intertwining matrices.   
 
An exactly analagous treatment of the higher dimensional theories
discussed above is possible. The only real difference is the replacement 
of $SU(2)$ by $SU(N)$, and the corresponding replacement of Gauss law
invariant states by $SU(N)$ spin-networks. The construction of
diffeomorphism invariant states remains unaltered. 

It is possible to couple matter fields to the action 
(\ref{N2}). Since an invertible metric is not available, the coupling 
is unusual. Nevertheless it has the property that a matter current appears 
with the spatial diffeormorphism constraint in the canonical theory in 
the expected way. 

Scalar field coupling  is achieved by adding to the action the 
term \cite{vh1}
\be 
      S[e,\phi,\pi] = \int_M d^{N^2}x\ \tilde{u}^a\pi\p_a\phi,  
\ee  
where $\tilde{u}^a$ is given by Eqn. (\ref{u}). The momentum conjugate 
to $\phi$ is $\tilde{\Pi}=\tilde{u}^0\pi$, and  
the the diffeomorphism constraint has the additional piece 
$\tilde{\Pi}\p_a\phi$. 

There is a similar coupling to a doublet $\psi^i$ and $\chi^i$ of 
$SU(N)$ matter fields, obtained by adding to the action (\ref{N2}) 
the term 
\be 
  S[A,e;\psi,\chi] = \int _M d^{N^2}x\ \tilde{u}^a k_{ij} \chi^iD_a\psi^j, 
\ee 
where $D$ is the covaraint derivative of $A$. The momentum conjugate to 
$\psi^i$ is $\tilde{P}^i_\psi = \tilde{u}^0\chi^i$, and now the Gauss law 
also acquires a source term $f^i_{\ jk}\tilde{P}_\psi^j \psi^k$. 

Finally, a coupling to ``Rarita-Schwinger''  $SU(N)$ fields 
$(\pi^i_a,\psi_a^i)$ is also possible, and is achieved with the term 
\be 
     S[e,A;\pi,\psi] = \int_M \eta_{i_1\cdots i_{N^2-1}} e^{i_1}\wedge 
\cdots \wedge e^{i_{N^2-3}}\wedge \pi^{i_{N^2-2}}\wedge 
D\wedge \psi^{i_{N^2-1}}. 
\ee 
This is similar to a coupling to two component spinors 
introduced for the $SU(2)$ theory \cite{vh2}. 

Quantization of the four-dimensional $SU(2)$ theory with matter couplings 
has been discussed \cite{T,BK}. For the case of scalar and $SU(N)$ 
spinors,  it is evident that a  similar approach will work in higher 
dimensions. For the Rarita-Schwinger type fields, however, a 
generalization would be necessary because of the spatial indices carried 
by these fields. 

So far we have discussed a theory similar to (\ref{su(2)}), but in higher dimensions, which does not have the Hamiltonian constraint in its 
canonical formulation. It is possible to add this constraint by hand to 
the Hamiltonian theory given above. It has the form  
\be 
 \tilde{e}^a_i\tilde{e}^b_j F_{ab}^k f_k^{\ ij} = 0, 
\label{H}
\ee
which is identical to that for gravity in four dimensions in the 
Ashtekar variables. The constraint algebra remains first class, and 
is identical to the four dimensional case. 
 
Although adding the constraint (\ref{H}) by hand does give a 
classically consistent system, there are in fact 
higher dimensional $SU(N)$ actions for which this can be achieved 
directly. Consider the action 
\be 
 S^{N^2-1} = \int_M \eta_{i_1\cdots i_{N^2-2}} 
e^{i_1}\wedge \cdots \wedge e^{i_{N^2-3}}\wedge F^{i_{N^2 -2}}(A).
\label{N2-1}
\ee
This is similar to (\ref{N2}), but with the differences that 
the Lagrangian density is an ($N^2-1$)--form and 
$\eta_{i_1\cdots i_{N2-2}}$ is an antisymmetric tensor of rank $N^2-2$. 
The metric is now no longer degenerate since there are ($N^2-1$) fields 
$e_a^i$, in the same number of dimensions. For $N=2$,  
$\eta_{i_1i_2} = f^{k_0}_{\ i_1 k_1} f^{k_1}_{\ i_2 k_0} = k_{i_1i_2}$ 
is the Cartan metric, and (\ref{N2-1}) is just the action of 
3-dimensional gravity.  

Being first order in derivatives, it is straightforward to construct 
the Hamiltonian version of this theory. The canonical theory does contain 
the time diffeomorphism constraint (\ref{H}). To see this vary (\ref{N2-1}) 
with respect to $e_0^i$ (the ``time'' component of $e_a^i$). This gives 
the $N^2-1$ constraints 
\be 
\eta_{ii_1\cdots i_{N^2-3}}\
\tilde{\epsilon}^{a_1\cdots a_{N^2-2} }\ 
 \left( e_{a_1}^{i_1} \cdots e_{a_{N^2-4}}^{i_{N^2-4}}\right)\  
F_{a_{N^2-3}a_{N^2-2}}^{i_{N^2-3}} = 0,
\label{C}
\ee
where all the world indices are spatial. This contains the spatial
diffeomorphism and Hamiltonian constraints. 
To extract these as functions of the canonical 
variables, note first that the momentum conjugate to the connection is 
\be
\tilde{e}^a_i = \eta_{ii_1\cdots i_{N^2-3}}\
\tilde{\epsilon}^{aa_1\cdots a_{N^2-3}}\ 
\left( e_{a_1}^{i_1}\cdots e_{a_{N^2-3}}^{i_{N^2-3}}\right).
\ee
Therefore, contracting (\ref{C}) with  $e_b^i$ (where the index $b$ is 
spatial) gives the $N^2-2$ constraints  
\be 
    \tilde{e}^a_i F_{ab}^i = 0.
\ee
Noting that $A_0^i$ is the lagrange multiplier for the Gauss law, 
the combination  (\ref{diff}) gives the spatial diffeomorphism 
constraint. To extract the Hamiltonian constraint in the form (\ref{H}), 
we must contract the free index $i$ in Eqn. (\ref{C}) to get a scalar 
density. This is achieved by contracting $(\ref{C})$ with 
\be 
\eta_{ij_1\cdots j_{N^2-2}} 
\tilde{\epsilon}^{b_1\cdots b_{N^2-2}}\ 
\left(e_{b_1}^{j_1}\cdots e_{b_{N^2-2}}^{j_{N^2-2}}\right)
\ ,
\ee
and simplifying. 

Finally, note that although the algebra of constraints of this $SU(N)$ 
theory is identical to that for four dimensional general relativity, 
it has more degrees of freedom than the latter. There are of course first 
order actions for general relativity of exactly the form (\ref{N2-1}), 
and therefore {\it with} a Hamiltonian constraint, for gauge group 
$SO(N-1,1)$. However their canonical formulations are no simpler than the  
usual metric variables due to the presence of second class 
 constraints\cite{abj}. The point of the $SU(N)$ action in the 
relevant dimension is that the canonical  theory has {\it exactly} the 
same constraints and algebra as the inherently four-dimensional Ashtekar 
variables.

In summary, we have made the following observations: (i) There are 
diffeomorphism invariant $SU(N)$ gauge theories in $N^2$ dimensions such 
that their Hamiltonian versions do not have a time reparametrization 
constraint, (ii) although the metric is degenerate, there is a prescription 
for matter couplings, (iii) full quantization along the lines of \cite{ash}
seems possible, and, (iv) In $N^2-1$ dimensions, there are $SU(N)$ theories  
which have the canonical constraint algebra of  general relativity, but   
have more degrees of freedom than the latter theory in the same dimensions.

\bigskip

\noindent {\it Acknowledgements:} This work was supported by the Natural 
Science and Engineering Research Council of Canada.

\end{document}